\PassOptionsToPackage{numbers}{natbib}
\documentclass[Afour,sageh,times]{sagej}

\usepackage{moreverb,url}
\usepackage{subfig}
\usepackage[export]{adjustbox}
\usepackage[colorlinks,bookmarksopen,bookmarksnumbered,citecolor=red,urlcolor=red]{hyperref}
\setcitestyle{numbers}

\newcommand\BibTeX{{\rmfamily B\kern-.05em \textsc{i\kern-.025em b}\kern-.08em
T\kern-.1667em\lower.7ex\hbox{E}\kern-.125emX}}

\def\volumeyear{2016}

\begin{document}

\runninghead{Smith and Wittkopf}

\title{Impacts of aspect ratio on task accuracy in parallel coordinates}

\author{Hugh Garner\affilnum{1}, Paul Burton\affilnum{2} and Sara Johansson Fernstad\affilnum{3}}

\affiliation{\affilnum{1} School of Computing, Newcastle University, UK.\\
\affilnum{2} Professor Emeritus of Data Science for Health, Newcastle University, UK\\
\affilnum{3} School of Computing, Newcastle University, UK\\
}

\corrauth{Hugh Garner, School of Computing
Urban Sciences Building,
1 Science Square,
Newcastle-upon-Tyne,
NE4 5TG, UK.}

\email{hugh.garner@newcastle.ac.uk}

\begin{abstract}
 Parallel coordinates plots (PCPs) are a widely used visualization method, particularly for exploratory analysis. Among the strengths of PCPs are their ability to support diverse tasks in multivariate data analysis, such as correlation estimation, value tracing and cluster identification. Previous studies show that PCPs perform much more poorly for estimating positive correlation than for estimating negative correlation \cite{Harrison2014-ks}, but it is not clear if this is affected by the aspect ratio (AR) of the axes pairs. In this paper, we present the results from an evaluation of the effect of the aspect ratio of axes in static (non-interactive) PCPs for two tasks: a) linear correlation estimation and b) value tracing. For both tasks we find strong evidence that aspect ratio influences accuracy, including aspect ratios greater than 1:1 being much more performant for estimation of positive correlations. We provide a set of recommendations for visualization designers using PCPs for correlation or value-tracing tasks, based on the data characteristics and expected use cases.
\end{abstract}

\keywords{parallel coordinates, aspect ratio, correlation}

\maketitle

\section{Introduction}\label{sec:introduction}
PCPs are valuable for multivariate data as they scale effectively with number of attributes displayed \cite{Munzner2014-pk,Inselberg2009-qg}, with increasing horizontal display space occupied as number of attributes increases (assuming vertical axes). PCPs are used in a diverse range of fields and applications particularly in multi-dimensional exploratory analysis  \cite{Heinrich2013-mm}. These tasks are comprised of a number of fundamental parts, notably trends, value-tracing (or outlier detection), correlation estimation and cluster identification \cite{Andrienko2001-ko,Munzner2014-pk,Shneiderman2003-nc,Dasgupta2010-vo,Heinrich2013-mm}. While there are numerous studies evaluating the performance of PCPs for these fundamental tasks (e.g. \cite{Harrison2014-ks,Li2010-fe}), little consideration is given to real-world use with differing numbers of axes and layout choices. In practice, the separation distance between PCP axes are frequently set simply by available horizontal screen space and the number of axes required for display \cite{Munzner2014-pk}, or by using a combination of metrics \cite{Dasgupta2010-vo}. There are few recommendations for minimum or maximum separation distance, or for aspect ratio (AR, the ratio of the horizontal axis separation distance to axis height). 

Prior work indicates that perception of patterns in scatter plots is significantly affected by scale and aspect ratio \cite{Cleveland1988-le} and as such it is reasonable to assume that PCPs may also be affected, leading to potentially variable performance dependent on choices made in spatial layout.

For example, Gehlenborg \& Wong \cite{Gehlenborg2012-ju} note that aspect ratio affects the perception of line crossings and suggest that axis height and separation distance should be adjusted such that the absolute value of all angles is close to 45 degrees.

In existing evaluations of PCPs, an aspect ratio of 1:1 (1) or 1:2 (0.5) is frequently used. For example, Li et al. \cite{Li2010-fe} compared the perception of correlation levels between scatter plots and PCPs, with both PCPs and scatter plots having a 1:1 ratio. 

To the best of our knowledge, there is no existing robust quantitative evaluation of the effect of aspect ratio on user tasks. It is plausible that differing aspect ratios may affect task success, especially given the differing patterns formed by positive and negative correlations and the associated performance difference \cite{Harrison2014-ks,Li2010-fe}. 

In parallel to our work, Meka et al. \cite{Meka2025-hy} conducted a study to explore simple negative/positive correlation perception in PCPs under varying aspect ratios and correlation coefficients. This study does not appear to find significant differences in accuracy of responses with the exception of 1:1 and 9:16. The response measure simply evaluated whether the participants correctly identified negative or positive correlations and as such inferences are very limited, with no consideration of the under- or over-estimation of the strength of a correlation.

This paper contributes three key elements to the field of visualization research:

\begin{itemize}
    \item Evaluation to determine whether aspect ratio has an effect on task success for value tracing and correlation estimation tasks in static (non-interactive) PCPs. Evidence shows that aspect ratio does impact on task success, with effect size dependent on both number of samples and aspect ratio.
    \item Provides evidence that correlation estimation is affected asymmetrically - that is, dependent on the sign of the correlation, with positive correlations being much more accurately estimated at aspect ratio $>$1.
    \item Recommendations for designers to set aspect ratio of axes dependent on expected primary tasks.
\end{itemize}

The remainder of the paper is structured as follows: Section \ref{sec:prior} presents related work on aspect ratio in visualizations. Hypotheses are described in Section \ref{sec:hypotheses}. The methodology used in our study is described in Section \ref{sec:methods}, followed by presentation of results in Section \ref{sec:results}. We discuss the results and limitations of the study in more detail in section \ref{sec:discussion}, followed by future directions in Section \ref{sec:future}.

\section{Prior work}\label{sec:prior}

While there exists extensive work that examines the effect of plot height or aspect ratio, it largely focuses on single channels or line plots \cite{Beattie2002-vf,Cleveland1984-se,Heer2009-js,Heer2010-mx}. Cleveland and McGill introduce the principle of "banking to 45 degrees" \cite{Cleveland1984-se} - ensuring that the average slope of a line in a line chart is 45 degrees - which has been frequently used as a guideline in optimization of aspect ratio. Based on this, a number of methods to optimise the aspect ratio of plots exist \cite{Fink2013-rl,Wang2018-fr}. However, this may not always be optimal or minimise error \cite{Talbot2012-qo,Heer2006-zm}.

\begin{figure*}[!t]
\centering
    \subfloat[]{\includegraphics[width=0.35\textwidth, valign=c]{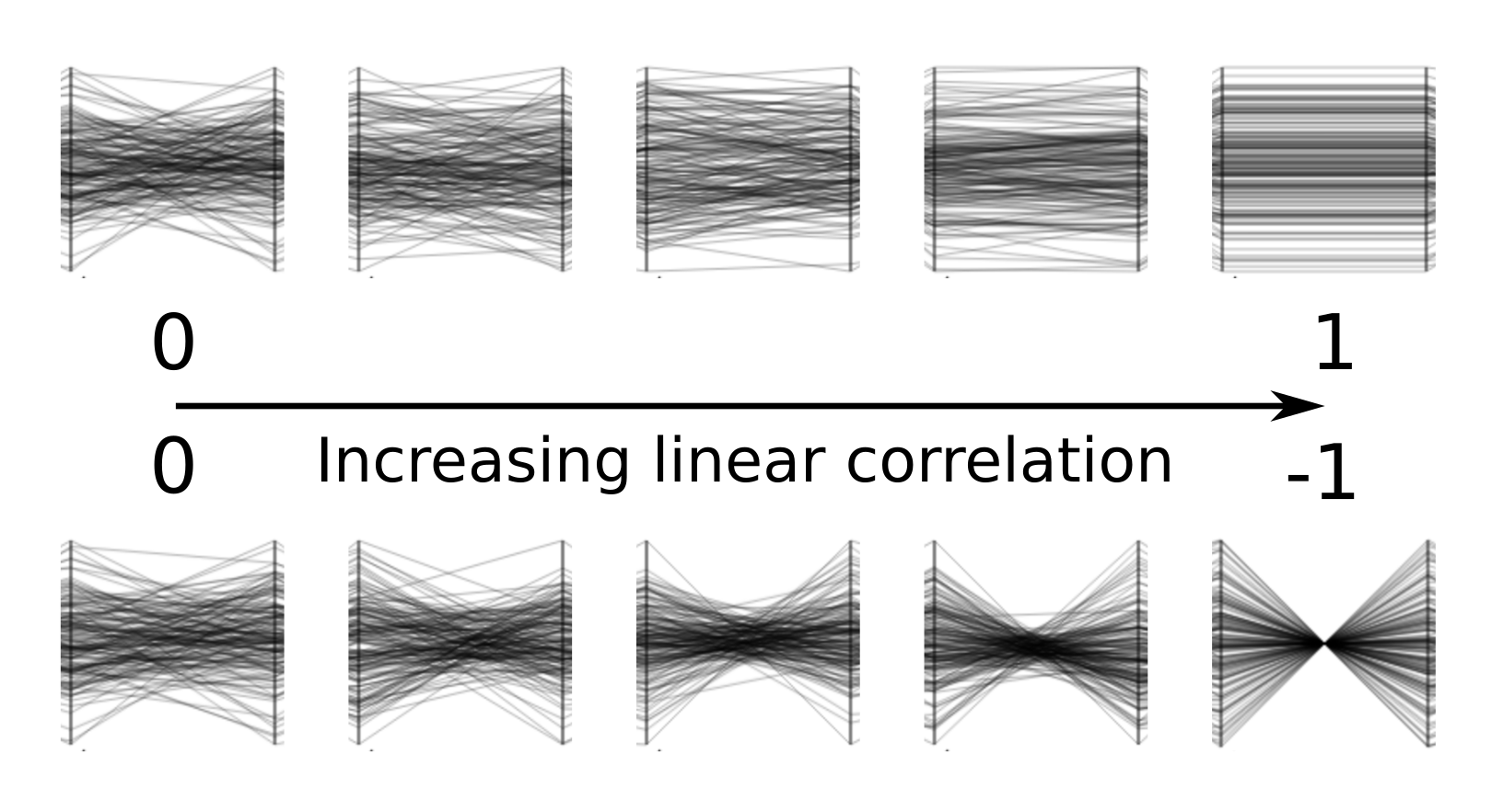}
    \label{fig:pcp_corr_examples}}
    \hfil
    \subfloat[]{\includegraphics[width=0.6\textwidth, valign=c]{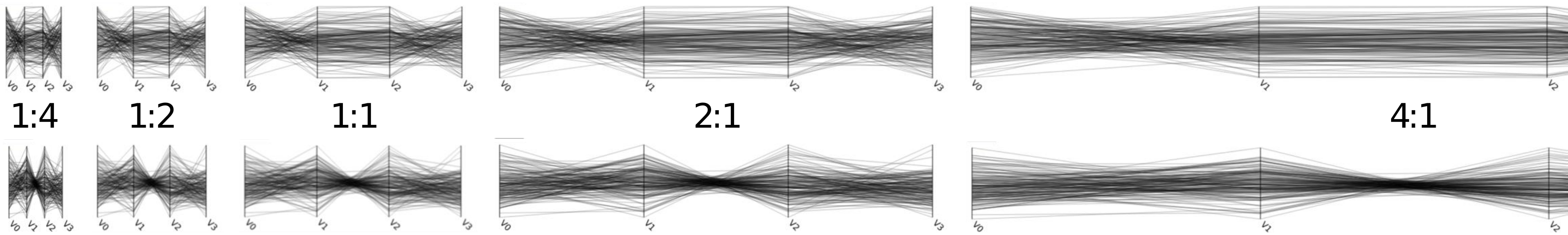}\label{fig:pcp_corr_aspect_ratio}}
    \caption{Correlation in a PCP: (a) patterns formed with increasing correlation strength; (b) patterns in a PCP by aspect ratio (central axes - top: positive correlation; bottom: negative correlation)}
    \label{fig:pcp_corr_detail}
\end{figure*}

Perception of correlation in scatter plots is affected by scale selection, with variables perceived as being more highly correlated when the scales are adjusted to make the point cloud tighter \cite{Cleveland1982-ne}. Linear correlations in a PCP form distinctive visual features dependent on sign (see fig. \ref{fig:pcp_corr_examples}), and correlation perception in a PCP is asymmetric, with negative correlations being perceived more distinctly than positive \cite{Harrison2014-ks}.

Dasgupta et al. \cite{Dasgupta2010-vo} introduce screen space metrics for PCPs. Given the limited resolution of a screen, they consider a number of metrics including line crossings, angle of crossings and parallelism that impact the geometric features on screen in relation to analysis tasks. These are used to optimise axis order on screen, with number of line crossings and parallelism found to be the most useful. The authors suggest that this is due to clutter reduction (line crossings contribute to clutter \cite{Ellis2006-ct}) and maximization of visibility of correlations. 

Pomerenke et al. \cite{Pomerenke2019-at} acknowledge the impact of the slope of polylines in a PCP for cluster identification tasks. Groups of near horizontal lines appear less clustered than groups where lines have a greater slope as line surface area and length increases with slope, leading to less apparent distance between lines. 

Decrease in relative line length with increasing aspect ratio may impact on slope estimation - Kennedy et al. \cite{Kennedy2006-nv} find that line length has an effect on angle estimation, with most accuracy when line pair length is approximately equal. 

Evidence suggests that aspect ratio has effects on measures other than slope. For example, Heer and Bostock \cite{Heer2010-mx} find that aspect ratio affects area estimation.

This provides a strong indication that features affected by aspect ratio will impact effectiveness of a visualization, but there is no work that directly evaluates the strength of any effect on a PCP.

\section{Hypotheses}\label{sec:hypotheses}
The aim of the research presented in this paper is to investigate whether there is a difference in task success at different axis aspect ratios in PCP. Given the specific aim of understanding task success, we do not propose perceptual mechanisms for such a difference. Nevertheless, there are some indications from studies of the effect of aspect ratio on other visualization methods that a difference may exist for the tasks we focus on (correlation estimation and value tracing).

Tasks were selected to represent common use-cases for a PCP in a exploratory visualization context \cite{Andrienko2001-ko}. Three tasks are commonly used in evaluations of a PCP - correlation estimation, value tracing and cluster identification \cite{Kanjanabose2015-ua}. Of these, we examine correlation estimation and value tracing as they have been extensively studied in PCPs previously (with single aspect ratios), providing a strongly validated experimental methodology, and are likely to be substantially affected by aspect ratio. Note that we vary aspect ratio by keeping the height of the PCP consistent, which is appropriate for real-world use cases where a PCP is stretched to fit the full width of the screen regardless of the number of axes. Varying the height of axes is also a potential use-case that may affect perception of patterns, but in order to limit the number of exposure variables it was considered beyond the scope of this initial investigation.

\subsection{Correlation estimation}

In a PCP, negative correlations form an ‘X’ shape, while lines have fewer crossings and become closer to parallel (assuming identical axis scales, i.e. a homogeneous graph) with increasing positive correlation (fig. \ref{fig:pcp_corr_examples}) \cite{Wegman1990-xy}. It is likely that distorting the plot horizontally will impact on the perception of these patterns, and, critically, the patterns may be affected differently. There is a negative relationship between aspect ratio and between-line angle - as horizontal distance is increased (while maintaining vertical distance), the angle between lines decreases. As such, it is likely that any component of the perceptual mechanism that uses angle estimation will be affected. Acute angles are often overestimated \cite{Howe2005-af} and the tilt effect \cite{Wenderoth1979-bt} is likely to apply with estimation of angles in context in a PCP, with bias strongest at around 20-30 degrees. Indeed, Akg\"oz et al. \cite{Akgoz2022-hp} find that angles between 0 and 10 deg have a strong attraction. Ware et al. \cite{Ware2002-em} suggest that acute angles are more likely to cause visual confusion than angles approaching 90 degrees, based on \cite{Blake1985-zr}. Line crossings will likely become less salient at small between-line angles, as is the case at high aspect ratios. For aspect ratios $>$ 1, strongly negative correlations may be less easily perceived (potentially mitigated by the clarity of the pattern). Fig. \ref{fig:pcp_corr_aspect_ratio} shows examples of high correlation values at different aspect ratios.
\hfill \break
\hfill \break
\textbf{Hypothesis 1:} 
\begin{itemize}
    \item \textbf{H1.1} Aspect ratio has an effect on accurate estimation of correlation in a PCP.
    \item \textbf{H1.2} Aspect ratio interacts with correlation to affect accuracy of estimated correlation.
    \item \textbf{H1.3} estimated correlation will more closely match the real value in PCPs with a high aspect ratio ($>$1) than in aspect ratios of 1 or lower.
\end{itemize}

\subsection{Tracing values}
PCPs are highly suited to tracing values across dimensions \cite{Kanjanabose2015-ua,Kuang2012-es}. As the horizontal distance between axes increases it is likely to be harder to follow a line between dimensions. Crossing angle has an effect on response time for path tracing tasks in graphs, decreasing as the angle increases to 70 degrees \cite{Huang2008-wc}. This effect is nonlinear, with a rapidly increasing response time as the angle drops below 20 degrees. For very small distances between axes the acute angle between the multiple polylines may cause visual clutter and decreased performance \cite{Johansson2016-rr}.
\hfill \break
\hfill \break
\textbf{Hypothesis 2:} 
\begin{itemize}
    \item \textbf{H2.1} Aspect ratio has an effect on successful tracing of values across multiple axes.
    \item \textbf{H2.2} as aspect ratio decreases below 1 or increases above 1 successful value tracing will become less likely.
\end{itemize}

\section{Methods}\label{sec:methods}
Two tasks were set to evaluate: a) whether aspect ratio has an effect on task accuracy and b) the relationship between aspect ratio, dataset size and task accuracy. 

\subsection{Procedures}

\paragraph{\textbf{Task A: Correlation estimation}}
Participants were asked to identify the correlation between a pair of adjacent PCP axes, by responding to whether the correlation was strongly, moderately, or weakly positive/negative, or if it was zero. 

\begin{figure*}[!t]
\centering

\subfloat[]{\includegraphics[width=0.5\textwidth]{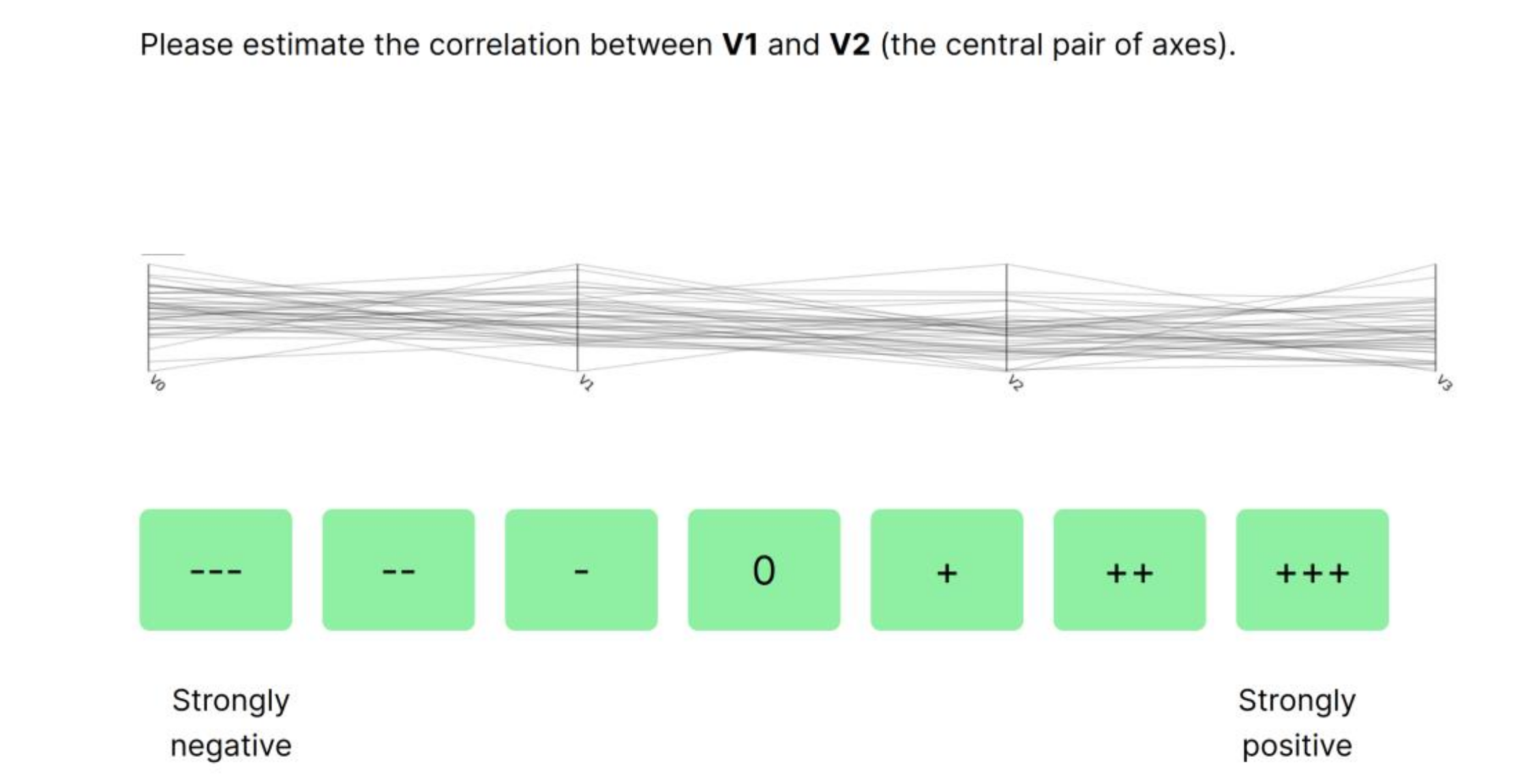}\label{fig_task_a_example}}
\hfil
\subfloat[]{\includegraphics[width=0.5\textwidth]{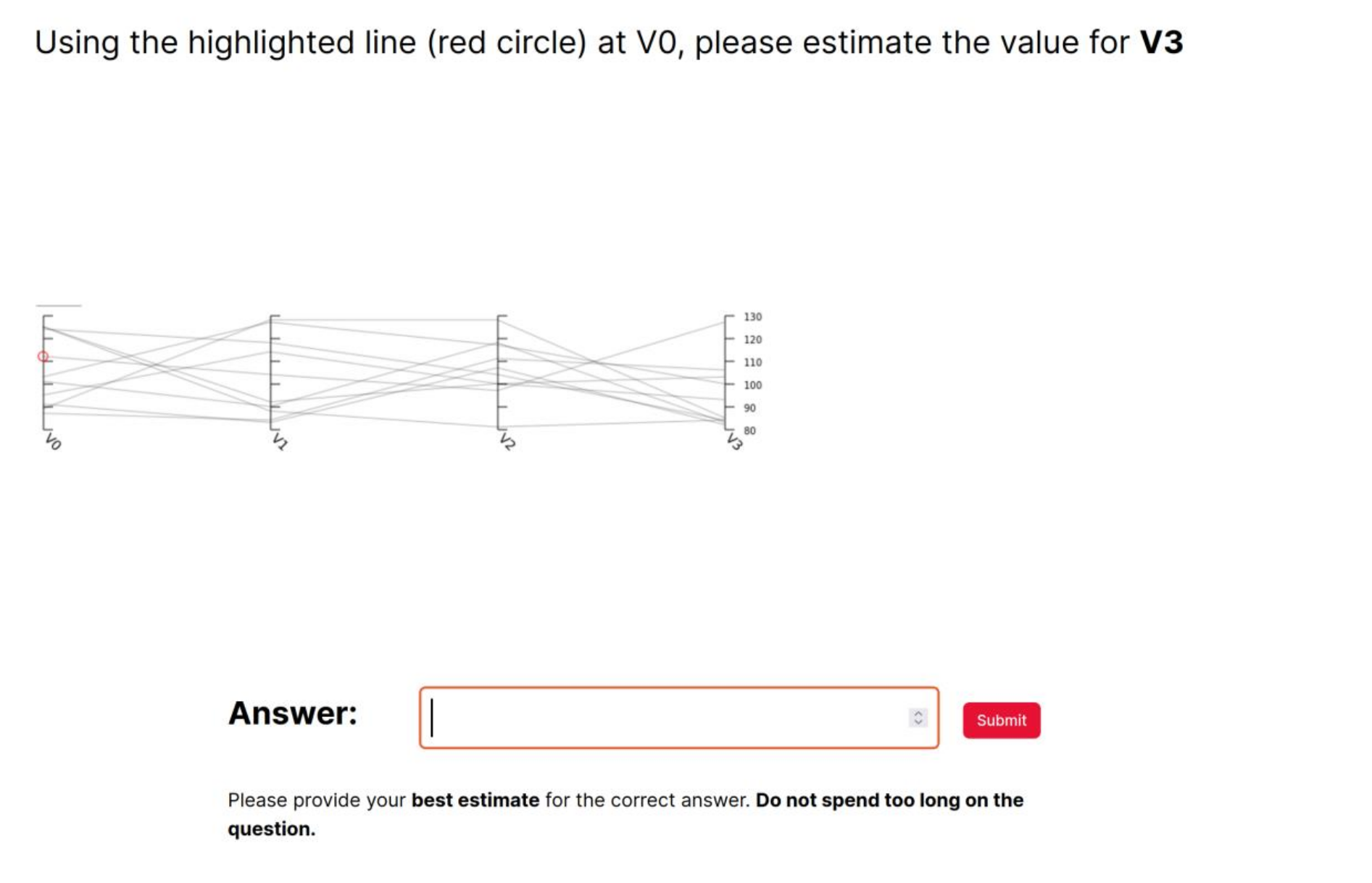}\label{fig_task_b_example}}
\caption{\label{fig:tasks_example}Example task screens (a) Task A: Correlation estimation. (b) Task B: Value tracing.}

\end{figure*}

The experimental setup was based on the study of correlation perception in PCPs and scatter plots by Li et al. \cite{Li2010-fe}. We adopted an identical approach for selecting correlation strengths, using equidistant levels on the theoretically optimal Fisher-z scale. The Fisher-z transformation transforms the sampling distribution of Pearson's r into a normal distribution, reducing skew at high values of r. Li et al. \cite{Li2010-fe} find that this theoretically optimal scale is a reasonable reflection of the human perceptual scale for PCPs. We note the relevant Pearson's r values associated with the selected Fisher-z scores in the parentheses below. Datasets of 4 variables with a random normal distribution were created for each sample size (40 and 160 samples) for each correlation between the question variables ($r$ = +/- 0.905 z = +/- 1.5; $r$ = +/- 0.762 z = +/- 1; $r$ = +/- 0.462 z = +/- 0.5; $r$ = 0 z = 0). Static images of the 4-axis PCP were generated for each size/correlation set and these were then scaled for each aspect ratio. Buttons were displayed below the plot, with a magnitude indicator for strong (+++/---), moderate (++/--), weak (+/-) or zero (0) correlation (see fig. \ref{fig_task_a_example}). Participants selected a button with the appropriate magnitude and sign of correlation as their response.

Aspect ratios were chosen to provide a realistic and readable plot on a single screen with 4 variables with a central aspect ratio of 1:1, using a log scale to ensure coverage at the ends of the distribution while closely simulating real values around the 1:1 ratio. Five aspect ratios at 0.25:1, 0.5:1, 1:1, 2:1 and 4:1 were then validated with a pilot study of 10 participants, which indicated that responses had appropriate range with minimal clustering at the extremes. To ensure data quality, task length was reduced by running the experiment in two parts, the first with aspect ratios of 0.25:1, 1:1 and 4:1 and the second with aspect ratios of 0.5:1, 1:1 and 2:1. Questions were randomised to mitigate learning effects where participants recognised the pattern in a plot scaled up or down. This was further mitigated by including the question order as a random effect in the analysis.

\paragraph{\textbf{Task B: Value tracing}}
Using a 4 variable PCP, users were asked to estimate the value in the last axis of the polyline circled at its intersection with the first axis.

Datasets of 4 variables with a random uniform distribution (range 0--50) were created for each sample size (10, 20, 40, 160). Following the procedure set out in \cite{Kuang2012-es}, to minimise confusion the initial and final values were specifically chosen to ensure no other sample values were closer than 3. Following the procedure in Task A, aspect ratios of 0.25:1, 0.5:1, 1:1, 2:1 and 4:1 were selected to characterise the dynamics of changing response with aspect ratio, then validated with a pilot of 10 participants which indicated reasonable variance of responses within aspect ratio levels. The pilot indicated that the dataset with 160 samples produced very few correct responses, probably due to visual clutter making the task unfeasible at the required screen resolutions. As such, the 160 sample datasets were removed from the final task.

For each sample size, 3 datasets were generated, creating 9 datasets in total. For each of these datasets, static images of PCPs were created for each aspect ratio, leading to a total of 45 plots/questions. A scale was printed on the final axis (the query axis), with a tick at intervals of 5.

Following a pilot study that indicated different datasets could create substantial variability in accuracy, we used the same dataset for each aspect ratio, generating 5 plots per dataset. To reduce the risk or a learning effect the answer scale was shifted by aspect ratio x 40 and question order was randomised on a per-participant basis.

\paragraph{\textbf{Task structure \& training}}
The tasks were implemented using the Gorilla online framework (gorilla.sc). Questions used static images of PCPs, with participants either selecting a button for correlation strength (Task A, fig. \ref{fig:tasks_example}a) or inputting the estimated value in a text entry field (Task B, fig. \ref{fig:tasks_example}b). Question order was randomised on a per-participant basis. For each task, participants were first given guidance on the task (for example correlation patterns in a PCP) and then completed a set of training questions. Training questions asked participants to estimate the correct answer and then click a button to reveal the correct answer. Training questions were formatted identically to the live questions, but with some additional guidance to ensure participant comprehension (Task A). Participants were free to spend as long as they wished completing the training questions. Once training questions were complete, the participants were asked to begin the live questions. Participants were asked 42 questions for Task A and 45 questions for Task B.

\paragraph{\textbf{Data quality}}Three ‘attention check’ questions were asked interspersed evenly between the trial questions. Where participants incorrectly answered one or more of these questions, all responses from that participant were removed. Responses under 850ms were removed based on tests of minimum response times and the distribution of real responses. For Task A, this resulted in the removal of 43 out of 8400 responses (involving 13 participants), resulting in 8298 responses for analysis. For Task B, no responses were removed, resulting in 2160 responses for analysis.

\paragraph{\textbf{Participants}}
Participants were recruited using the online platform Prolific (prolific.co) and were screened to ensure normal or corrected-normal vision. Participants were asked for consent then asked to rate their experience with visualization  (scale 0-5) and statistics (scale 0-5). Demographic data and responses for these questions are detailed in the supplementary material. Participants were paid \pounds1.60 for participation in Task A and \pounds3 for participation in Task B. To prevent learning effects, participants that completed the first part of Task A (with aspect ratios 0.25, 1 and 4) were not eligible to complete the second part. For Task A, 205 participants (105 for part 1 and 100 for part 2) were recruited with 200 remaining after quality control (QC) (103 for part 1 and 97 for part 2); for Task B 49 participants were recruited with 48 remaining after QC. 

Participants were asked their age range for both tasks: for Task A the majority of participants were under 40 (86\%); for Task B 87.6\% were under 40. For further details, see supplementary materials. Participants were also asked a series of questions asking for details of background knowledge of statistics and visualization. For Task A, participants were asked to a rate on a Likert scale (1-5) a) their knowledge and experience of visualization methods (median response 3, 75.5\% 3 or lower) and b) their knowledge and experience of statistical methods and analysis (median response 3, 74\% 3 or lower).

For Task B, participants were asked for their years of experience in a) using and/or knowledge of data visualization methods (median 0-2 years, 77.1\% less than 2 years) and b) using and/or knowledge of statistical methods and analysis (median 0-2 years, 70.8\% less than 2 years). Participants were also asked for their experience of using Parallel Coordinates, with 72.9\% having no experience.

\subsection{Analysis methodology}

\paragraph{\textbf{Response measures}}
\textbf{Task A:} Accuracy was measured as the distance between the estimated and real correlation level. For example, a true Fisher-z value of -1.5 with a response of -1 ('--' in \ref{fig_task_a_example}) produces an accuracy measure of 0.5 - i.e. the participant slightly underestimated the absolute magnitude of the strongly negative correlation. Similarly, for a true Fisher-z value of 0.5 with a response of 1 ('++' in \ref{fig_task_a_example}) produces an accuracy measure of 0.5 - i.e. the participant slightly overestimated the absolute magnitude of the weakly positive correlation. For the simple test of correct responses, we also convert this to a binary outcome (if the estimated correlation level was within 1 category of the real level). 
\textbf{Task B:} Responses were measured as a binary outcome (correct or incorrect). To account for error of visual estimation in the response, responses were considered correct if they were within 2 (0-50 scale) of the correct value. Note that a continuous error scale (rather than binary correct/incorrect) cannot be used here as the most likely source of error is tracing an incorrect line, and any resultant error will depend on the endpoint of that line (which may be at any point on the scale).

\paragraph{\textbf{Statistical tests}}
The experiments are repeated measures (within subject) designs, with non-normal response distributions. To determine how likely the difference between factors was due to random chance we use either a Skillings-Mack test \cite{Skillings1981-my,Chatfield2009-yp} (Task A, incomplete block design) or Friedman test (Task B). Skillings-Mack is required for Task A as the experiment was performed in two parts so data is missing for moderate or strong correlation levels (part 1 and 2 respectively). Note that a Skillings-Mack test is exactly equivalent to a Friedman test where there is no missing data in a balanced complete-block design \cite{Chatfield2009-yp}.

Generalised linear mixed models (GLMMs) were used for analysis of the relationship between the aspect ratio, dataset size and accuracy. These allow the inclusion of random effects to account for the repeated measures within participants. For Task A, where accuracy is a measure of distance between the real and estimated correlation value, an ordered beta GLMM \cite{Kubinec2023-de} (logit link) was used. This is the most appropriate fit for the distribution of the response variable, which is on a bounded scale. The model enables evaluation of direction and magnitude of response and as such can indicate whether responses are likely to be under- or over- estimated. For Task B (accuracy as binary outcome), a binomial GLMM (logit link) was used. Models were fitted with interaction between aspect ratio, sample size (both tasks) and Fisher-z (Task A only) - we expect that sample size will affect perception of the differing patterns formed by positive or negative correlations and that the effect of aspect ratio will be dependent on sample size and correlation. To account for variation in performance between participants, a random effect for participant (intercept only) was included. Further, given the use of the same dataset for all levels of aspect ratio, we include a trial number variable as a random effect to account for potential learning effects of question order. See table \ref{tab:analysis_models} for model details.

\begin{table*}[t] \centering
    \renewcommand{\arraystretch}{1.5} \caption{Analysis models}
    \label{tab:analysis_models}
    \begin{tabular}{l l}
        \hline
        \textbf{Task} & \textbf{Model} \\
        \hline
        Task A & $
        response\_accuracy \sim fisher\_z*aspect\_ratio*sample\_size + (1|participant\_id) + (1|trial\_num)    
        $ \\
        Task B & $
        response\_correct \sim aspect\_ratio*sample\_size + (1|participant\_id) + (1|trial\_num)    
        $ \\
        \hline
    \end{tabular}
    \par\smallskip\noindent{\small Fitted models for analysis of Task A (top) and Task B (bottom), using the R notation (see \cite{Bates2015-jg}).}
\end{table*}

GLMMs were evaluated using AIC \cite{Akaike2011-dm} for relative comparisons of models with and without interactions, and residuals were checked using DHARMa \cite{Hartig2021-aw}. Models with complex higher-order interactions can be difficult to interpret directly, as coefficients in non-linear models with interactions do not map straightforwardly to changes on the response scale. We therefore follow standard practice \cite{Arel-Bundock2024-bt} by evaluating marginal predictions at the population level (fixed effects only, with random effects held at their expected value of zero), representing the expected response for an average participant in the study. We generate adjusted predicted responses for each level of the independent variables (e.g. sample size = 40 or 160, aspect ratio = 0.25, 0.5, etc, Fisher-z = -1.5, -1, etc) with the ‘ggpredict’ \cite{strengejacke-cv} function from the ggeffects \cite{Ludecke2018-hg} package. This provides population-level predictions from the full model ensuring that interaction effects are included in the estimates.

To evaluate the hypothesised interaction between aspect ratio and accuracy (H1.2), we formalise inferences from the fitted model for Task A with a measure comparing the variance of predicted responses. At each sample size (40 and 160), we fit a simple linear model to the predicted response data for each Fisher-z score, weighting by the inverse of the variance of the predicted value (link scale). Taking each pair of Fisher-z values, we then compare the variance of the residuals with an F-test to indicate whether the differences in variances are due to chance i.e. does the 'spread' of results by aspect ratio differ dependent on the true Fisher-z?

\section{Results}\label{sec:results}
\subsection{Task A}

\begin{figure}[htb]
\centering
\includegraphics[width=0.5\textwidth]{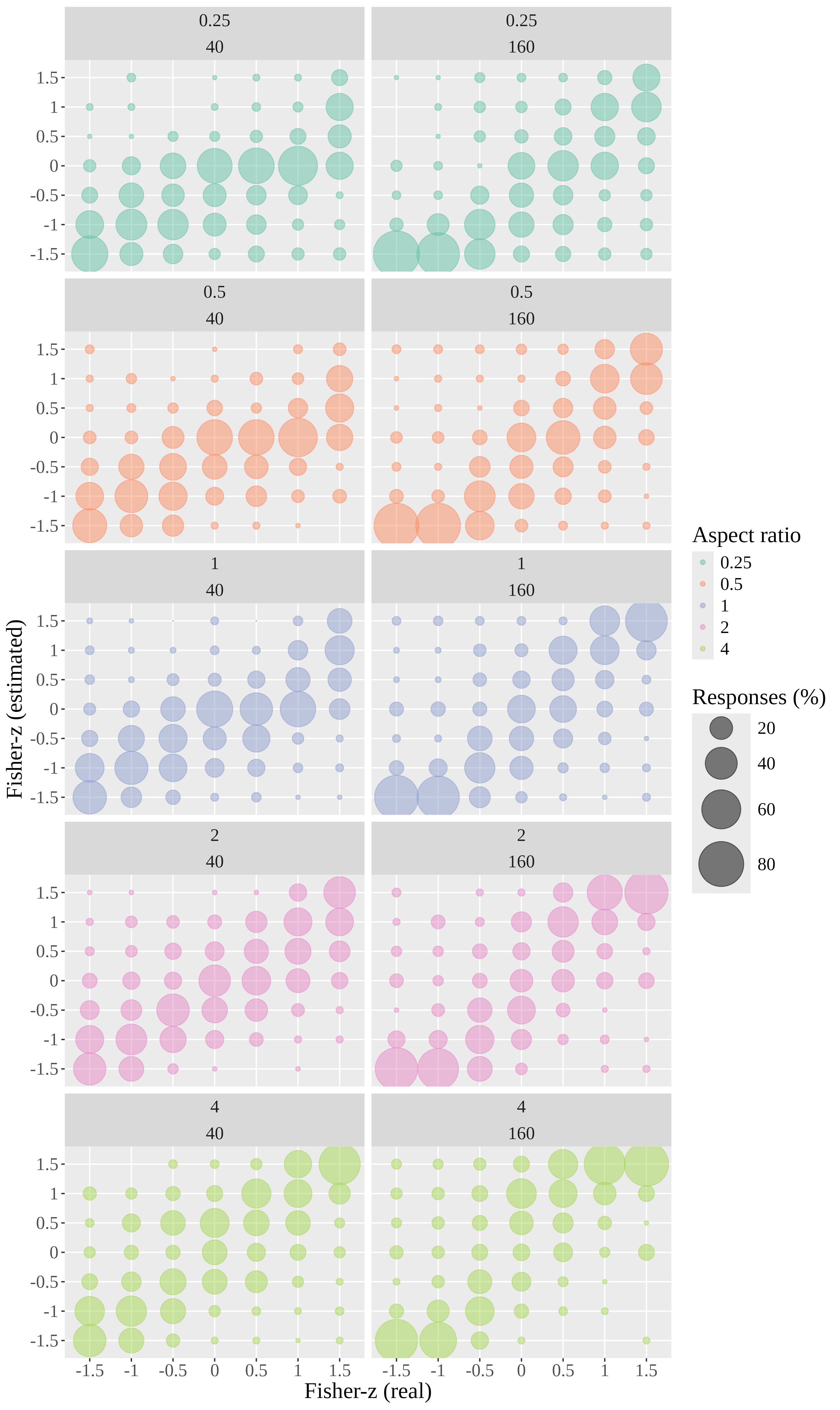}
\caption{Task A: Real (true) and estimated Fisher-z scores by aspect ratio (facet rows and colour) and sample size (40 and 160, facet columns). Each facet shows a combination of aspect ratio and sample size, and a facet then includes the true Fisher-z (x-axis) and participant estimated Fisher-z (y-axis). Bubble area indicates proportion of responses at each estimated Fisher-z value for the specific true Fisher-z value. }
\label{fig:corr_results_bubble}
\end{figure}

\textbf{H1.1: Aspect ratio has an effect on accurate estimation of correlation in a PCP.}
\newline
Fig. \ref{fig:corr_results_bubble} shows response percentage distributions for true Fisher-z by Fisher-z as estimated by participants. Values on the diagonal are correct, with values below and right of the diagonal being under-estimation and values above and left being over-estimation. There are distinct patterns for each aspect ratio, indicating that there is a high likelihood of a real difference in proportion of correct responses dependent on aspect ratio and providing support for \textbf{H1.1}. Skillings-Mack test results for sample size \textbf{40 stat. 59.9 $p$ 3.1 $\times$ 10\textsuperscript{-12}} and \textbf{160: stat. 23.6 $p$ 9.6 x 10\textsuperscript{-5}} indicate that these differences are unlikely to be due to chance.

Marginal predictions from the model are displayed in fig. \ref{fig:corr_ob_accuracy} (see supplementary for table of values). These represent population-level predictions evaluated at specific factor levels (e.g., sample size = 40 or 160, aspect ratio, Fisher-z), with other covariates held at representative values. Confidence intervals indicate uncertainty in the estimation of the fixed effects. This indicates the expected population mean error in response by sample size group at each aspect ratio and level of Fisher-z score tested. Observed (unadjusted) sample means computed directly from the raw (post-QC) responses (no random-effects adjustment) are included as a sanity check to compare with population-level model predictions.

The error between estimated and actual correlation (figure \ref{fig:corr_ob_accuracy}) indicates direction of mean expected response and is dependent on the real correlation, with negative errors indicating overestimation for negative correlations and underestimation for positive ones. For example, for a response distance of 0.5 a Fisher-z of -1 ($r$ of -0.762) is perceived as a Fisher-z of -0.5 ($r$ of -0.462); a Fisher-z of 1 ($r$ of 0.762) is perceived as 1.5 ($r$ of 0.905).

As such, figure \ref{fig:corr_ob_accuracy} indicates responses are slight under-estimates for both negative and positive correlations (i.e. correlation is estimated as being slightly less strong than the real value). This effect is asymmetric and there is clear divergence in response by aspect ratio - as aspect ratio decreases, responses are more likely to be substantially under-estimated for positive correlations. However, the effect varies by sample size. 

For a real Fisher-z of 1, the error of estimation decreases as aspect ratio increases - for sample size 40, AR 0.25 is likely to lead to underestimating the true value by over 1 - users estimate Fisher-z values of 1 ($r$ of 0.762) around 0 (-0.09). At AR 4 however, the expected mean underestimation is much smaller at 0.22 - so users estimate values of 1 at 0.78 ($r$ of 0.653).

\textbf{H1.2: Aspect ratio interacts with correlation to affect accuracy of estimated correlation.}
\newline
Fig. \ref{fig:corr_results_bubble} and fig. \ref{fig:corr_ob_accuracy} show the ordering of aspect ratios at each correlation level is largely consistent, but variance is not consistent across correlation levels (Fisher-z).

For each real Fisher-z score (x-axis) the responses have varying bounds (for example, at the lowest Fisher-z, the error in responses can only ever indicate a Fisher-z closer to zero, with a positive distance to the true value). We therefore consider the spread of expected error (distance to true value) by aspect ratio at each level of real Fisher-z. This spread appears to increase with Fisher-z (with the exception of Fisher-z 1.5 at sample size 160), suggesting that aspect ratio interacts with Fisher-z and supporting \textbf{H1.2}. Considering the top plot in figure \ref{fig:corr_ob_accuracy} (sample size of 40), for a true Fisher-z of -1.5, the confidence intervals of the predicted mean are overlapping for all aspect ratios and so there is no evidence for a real difference in mean, and this is consistent across both sample sizes. Conversely, at a true Fisher-z of 1 ($r$ of 0.762) there is little overlap of confidence intervals across aspect ratios, providing strong evidence for a real difference in the true means.

This gradual divergence of expected means and non-overlapping CIs as the Fisher-z increases reflects the Skillings-Mack test results. To further evaluate the interaction between aspect ratio and correlation estimation accuracy, we conducted F-tests of pairwise variance differences of the predicted response error for each real Fisher-z value (see section \ref{sec:methods} for details). Table \ref{tab:task_a_f_test} presents the results of these F-tests, highlighting significant differences in variance across different levels of Fisher-z. For a sample size of 40, there is strong evidence that differences between strongly negative and strongly positive correlations are unlikely to be due to chance. The same effect appears also to be present, though more weakly so, for a sample size of 160.

Evidence of interaction between aspect ratio and sample size is apparent in this differing response variability in figure \ref{fig:corr_ob_accuracy}.

\textbf{H1.3: estimated correlation will more closely match the real value in PCPs with a high aspect ratio ($>$1) than in aspect ratios of 1 or lower.}
\newline
For a sample size of 40, responses appear likely to be closer estimates of the real value when aspect ratio exceeds 1. Specifically, aspect ratios of 2 and 4 show reduced estimation error, particularly for positive correlations. Further, there is a clear falloff in accuracy for AR $\leq$ 1 where correlation is greater than zero. This provides partial support for \textbf{H1.3}. Aspect ratios 0.25 and 0.5 have substantial under-estimation of positive correlations, but the drop in performance for AR 1 at correlation $>$ 0 is much less evident. There appears to be little evidence for a difference in performance for AR $\geq$ 1 at the highest correlation (Fisher-z of 1.5), with a substantial overlap in confidence intervals (AR 1 -0.33 CI95 -0.43 to -0.24; AR 2 -0.27 CI95 -0.40 to -0.13; AR 4 -0.21 CI95 -0.34 to -0.08). Responses for AR 4 indicate large overestimates of positive correlation, and notably zero correlation tends to be estimated as positive (0.42 CI95 0.29 - 0.55).

Participant is a substantial random effect (variance logOR 0.71), with the pooled model likely to be insufficient (Chisq $p$ 2.2 x 10\textsuperscript{-16}). Trial number (i.e. question order within factors) is insignificant as a random effect.

\begin{table}[htb]
\centering
\caption{Task A: Pairwise F-tests of predicted variance by sample size and real Fisher-z. Direction indicates which of the Fisher-z 1 or Fisher-z 2 values were greater. Sig. column indicates P-values under the thresholds of $5 \times 10^{-2}$ (\textbf{*}), $5 \times 10^{-3}$ (\textbf{**}), $5 \times 10^{-4}$ (\textbf{***}). }
\label{tab:task_a_f_test}
\resizebox{\columnwidth}{!}{\begin{tabular}{ccrrrl}
\toprule
Fisher-z 1 & Fisher-z 2 & F stat & P-value & Sig. & Direction \\ 
\midrule\addlinespace[2.5pt]
\multicolumn{6}{l}{Sample size: 40} \\ 
\midrule\addlinespace[2.5pt]
-1.5 & -1 & $2.7$ & $1.8 \times 10^{-1}$ & \textbf{} & V1 $>$ V2 \\ 
-1.5 & -0.5 & $5.9$ & $5.6 \times 10^{-2}$ & \textbf{} & V2 $>$ V1 \\ 
-1.5 & 0 & $4.3$ & $9.3 \times 10^{-2}$ & \textbf{} & V2 $>$ V1 \\ 
-1.5 & 0.5 & $13.8$ & $1.3 \times 10^{-2}$ & \textbf{*} & V2 $>$ V1 \\ 
-1.5 & 1 & $21.8$ & $5.6 \times 10^{-3}$ & \textbf{*} & V2 $>$ V1 \\ 
-1.5 & 1.5 & $16.4$ & $9.6 \times 10^{-3}$ & \textbf{*} & V2 $>$ V1 \\ 
-1 & -0.5 & $15.8$ & $1.0 \times 10^{-2}$ & \textbf{*} & V2 $>$ V1 \\ 
-1 & 0 & $11.5$ & $1.8 \times 10^{-2}$ & \textbf{*} & V2 $>$ V1 \\ 
-1 & 0.5 & $36.8$ & $2.1 \times 10^{-3}$ & \textbf{**} & V2 $>$ V1 \\ 
-1 & 1 & $58.0$ & $8.5 \times 10^{-4}$ & \textbf{**} & V2 $>$ V1 \\ 
-1 & 1.5 & $43.6$ & $1.5 \times 10^{-3}$ & \textbf{**} & V2 $>$ V1 \\ 
-0.5 & 0 & $1.4$ & $3.8 \times 10^{-1}$ & \textbf{} & V1 $>$ V2 \\ 
-0.5 & 0.5 & $2.3$ & $2.2 \times 10^{-1}$ & \textbf{} & V2 $>$ V1 \\ 
-0.5 & 1 & $3.7$ & $1.2 \times 10^{-1}$ & \textbf{} & V2 $>$ V1 \\ 
-0.5 & 1.5 & $2.8$ & $1.7 \times 10^{-1}$ & \textbf{} & V2 $>$ V1 \\ 
0 & 0.5 & $3.2$ & $1.4 \times 10^{-1}$ & \textbf{} & V2 $>$ V1 \\ 
0 & 1 & $5.1$ & $7.3 \times 10^{-2}$ & \textbf{} & V2 $>$ V1 \\ 
0 & 1.5 & $3.8$ & $1.1 \times 10^{-1}$ & \textbf{} & V2 $>$ V1 \\ 
0.5 & 1 & $1.6$ & $3.3 \times 10^{-1}$ & \textbf{} & V2 $>$ V1 \\ 
0.5 & 1.5 & $1.2$ & $4.4 \times 10^{-1}$ & \textbf{} & V2 $>$ V1 \\ 
1 & 1.5 & $1.3$ & $3.9 \times 10^{-1}$ & \textbf{} & V1 $>$ V2 \\ 
\midrule\addlinespace[2.5pt]
\multicolumn{6}{l}{Sample size: 160} \\ 
\midrule\addlinespace[2.5pt]
-1.5 & -1 & $1.7$ & $3.1 \times 10^{-1}$ & \textbf{} & V2 $>$ V1 \\ 
-1.5 & -0.5 & $2.8$ & $1.7 \times 10^{-1}$ & \textbf{} & V2 $>$ V1 \\ 
-1.5 & 0 & $7.5$ & $3.9 \times 10^{-2}$ & \textbf{*} & V2 $>$ V1 \\ 
-1.5 & 0.5 & $12.5$ & $1.6 \times 10^{-2}$ & \textbf{*} & V2 $>$ V1 \\ 
-1.5 & 1 & $10.2$ & $2.3 \times 10^{-2}$ & \textbf{*} & V2 $>$ V1 \\ 
-1.5 & 1.5 & $3.7$ & $1.2 \times 10^{-1}$ & \textbf{} & V2 $>$ V1 \\ 
-1 & -0.5 & $1.6$ & $3.2 \times 10^{-1}$ & \textbf{} & V2 $>$ V1 \\ 
-1 & 0 & $4.3$ & $9.3 \times 10^{-2}$ & \textbf{} & V2 $>$ V1 \\ 
-1 & 0.5 & $7.2$ & $4.1 \times 10^{-2}$ & \textbf{*} & V2 $>$ V1 \\ 
-1 & 1 & $5.9$ & $5.7 \times 10^{-2}$ & \textbf{} & V2 $>$ V1 \\ 
-1 & 1.5 & $2.2$ & $2.4 \times 10^{-1}$ & \textbf{} & V2 $>$ V1 \\ 
-0.5 & 0 & $2.6$ & $1.9 \times 10^{-1}$ & \textbf{} & V2 $>$ V1 \\ 
-0.5 & 0.5 & $4.4$ & $9.0 \times 10^{-2}$ & \textbf{} & V2 $>$ V1 \\ 
-0.5 & 1 & $3.6$ & $1.2 \times 10^{-1}$ & \textbf{} & V2 $>$ V1 \\ 
-0.5 & 1.5 & $1.3$ & $4.0 \times 10^{-1}$ & \textbf{} & V2 $>$ V1 \\ 
0 & 0.5 & $1.7$ & $3.2 \times 10^{-1}$ & \textbf{} & V2 $>$ V1 \\ 
0 & 1 & $1.4$ & $3.9 \times 10^{-1}$ & \textbf{} & V2 $>$ V1 \\ 
0 & 1.5 & $2.0$ & $2.6 \times 10^{-1}$ & \textbf{} & V1 $>$ V2 \\ 
0.5 & 1 & $1.2$ & $4.2 \times 10^{-1}$ & \textbf{} & V1 $>$ V2 \\ 
0.5 & 1.5 & $3.4$ & $1.3 \times 10^{-1}$ & \textbf{} & V1 $>$ V2 \\ 
1 & 1.5 & $2.7$ & $1.8 \times 10^{-1}$ & \textbf{} & V1 $>$ V2 \\ 
\bottomrule
\end{tabular}

 }
\end{table}

\begin{figure*}[htb]
\centering
    \includegraphics[width=1\textwidth]{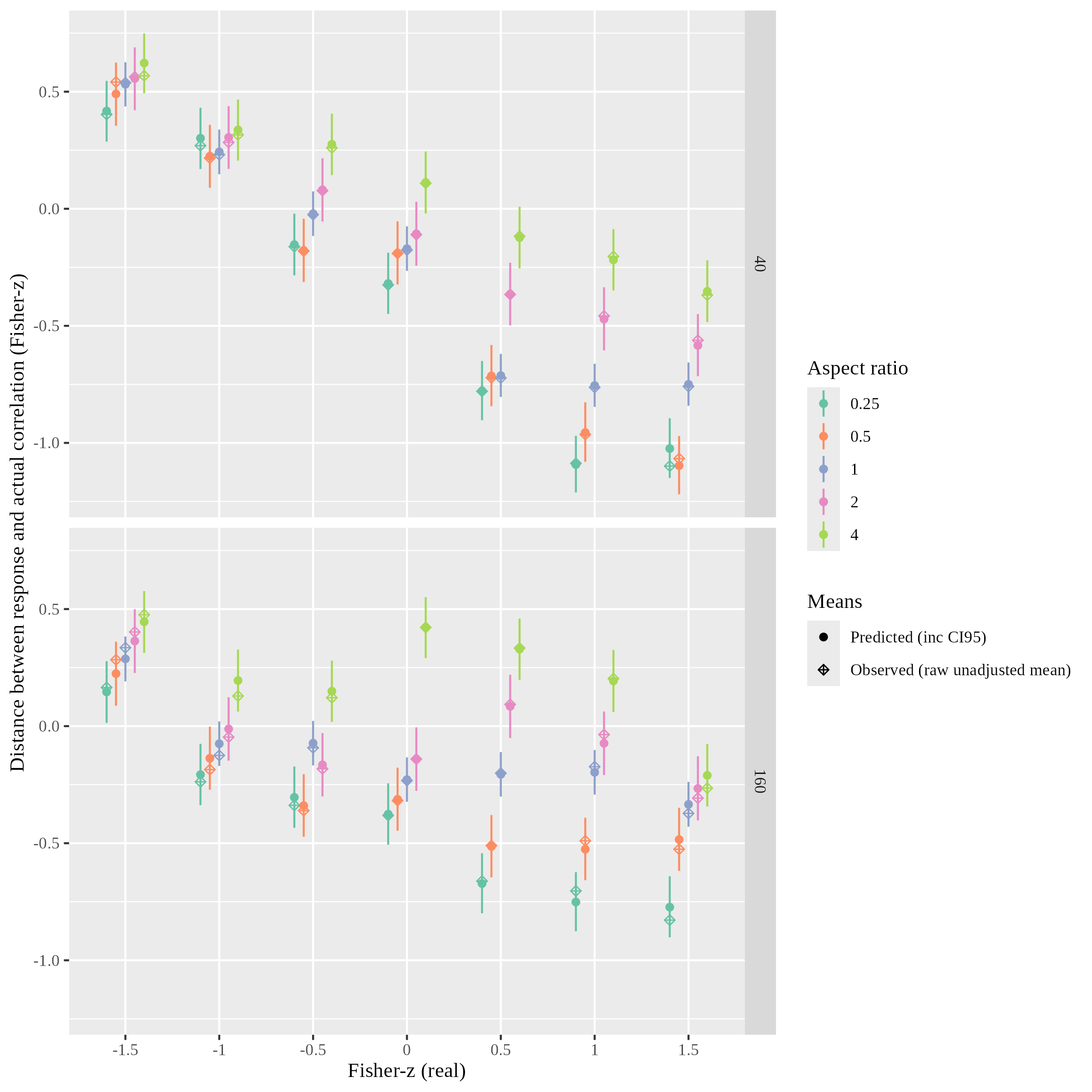}
\caption{Task A: Marginal (population-level) model predictions of signed error (y-axis, Fisher-z) by true Fisher-z (x-axis) across aspect ratios (colour) and sample sizes (top facet sample size 40, bottom 160), with 95\% CIs, plotted together with the corresponding observed (unadjusted) mean errors. For any given true Fisher-z value (x-axis), the distance of the error (y-axis) indicates the magnitude of the mean error in estimated value by participants. A positive value indicates over-estimation, a negative under-estimation.} 
\label{fig:corr_ob_accuracy}
\end{figure*}

\subsection{Task B}

\textbf{H2.1: Aspect ratio has an effect on successful tracing of values across multiple axes.}
\newline
Friedman tests for number of correct responses by aspect ratio at each sample size (sample size \textbf{10: $p$ 3.4 x 10\textsuperscript{-15}}; \textbf{20: $p$ 1.1 x 10\textsuperscript{-13}}; \textbf{40: $p$ 1.5 x 10\textsuperscript{-12}}) provide strong evidence that aspect ratio has an effect on successful tracing of values across axes (\textbf{H2.1}). 

\begin{figure*}[htb]
\centering
    \includegraphics[width=1\textwidth]{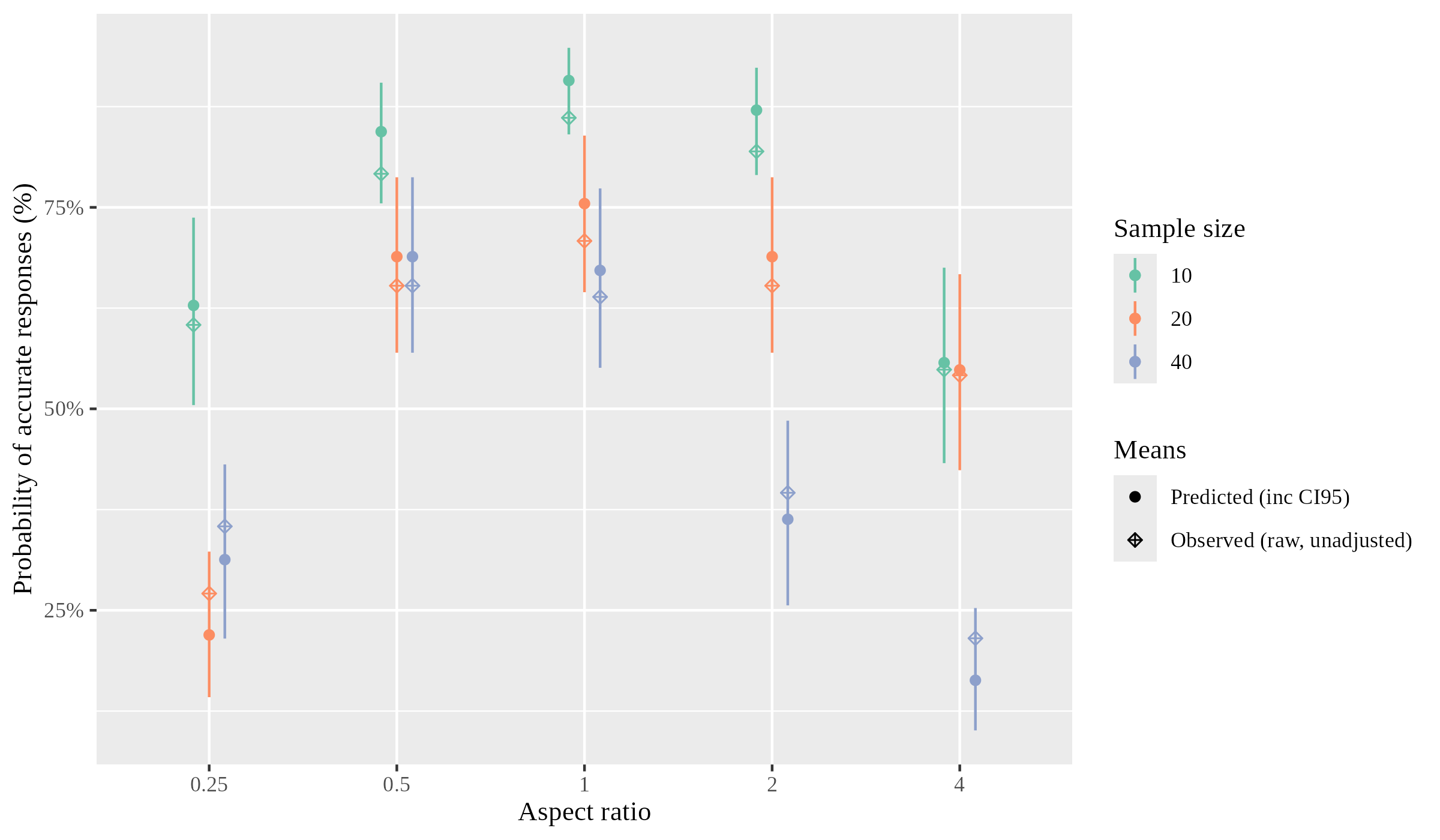}
\caption{\label{fig:vt_accuracy_predicted}Task B: Marginal effects - model-based marginal predictions and 'real' (observed, unadjusted) probability correct by aspect ratio and sample size (number of polylines)}
\end{figure*}

\textbf{H2.2: as aspect ratio decreases below 1 or increases above 1 successful value tracing will become less likely.}
\newline
Fig. \ref{fig:vt_accuracy_predicted} shows marginal probability of an accurate response by aspect ratio and sample size for the population (see supplementary for table of values). Predictions are marginal, evaluated at specific factor levels (e.g., sample size = 10, 40 or 160, aspect ratio), with other covariates held at representative values. For all sample sizes, AR 1 is most accurate. There is strong evidence for a substantial reduction in accuracy at AR 0.25 (size 10 OR 0.17; size 20 OR 0.09; size 40 OR 0.22) and 4 (size 10 OR 0.13; size 20 OR 0.39; size 40 OR 0.1) which are unlikely to be due to chance (see table \ref{tab:task_b_t_test}), supporting \textbf{H2.2}. There is some very limited evidence for a reduction in accuracy at AR 0.5 and 1 for sample size 10 (OR 0.55). The expected probability of accurate responses does appear to interact with both aspect ratio and sample size. There is good evidence for a substantial reduction in accuracy at AR 2 compared to AR 1 for sample size 40 (OR 0.28, $p$ 5.3 x 10\textsuperscript{-4}), but no apparent reduction at AR 0.5 (OR 1.08). This asymmetry is not apparent for sample sizes 10 or 20.

\begin{table}[htb]
\centering
\caption{Task B: Pairwise evaluation of estimated accuracy by aspect ratio. Sig. column indicates P-values under the thresholds of $5 \times 10^{-2}$ (\textbf{*}), $5 \times 10^{-3}$ (\textbf{**}), $5 \times 10^{-4}$ (\textbf{***}).}\label{tab:task_b_t_test}
\clearpage{}\begin{tabular}{cccrr}
\toprule
Sample size & AR 1 & AR 2 & P-value & Sig. \\ 
\midrule\addlinespace[2.5pt]
10 & 0.25 & 0.5 & $1.5 \times 10^{-3}$ & ** \\ 
10 & 0.25 & 1 & $1.5 \times 10^{-6}$ & *** \\ 
10 & 0.25 & 2 & $1.6 \times 10^{-4}$ & *** \\ 
10 & 0.25 & 4 & $4.2 \times 10^{-1}$ &  \\ 
10 & 0.5 & 1 & $1.4 \times 10^{-1}$ &  \\ 
10 & 0.5 & 2 & $5.9 \times 10^{-1}$ &  \\ 
10 & 0.5 & 4 & $3.3 \times 10^{-4}$ & *** \\ 
10 & 1 & 2 & $4.0 \times 10^{-1}$ &  \\ 
10 & 1 & 4 & $4.5 \times 10^{-6}$ & *** \\ 
10 & 2 & 4 & $6.3 \times 10^{-5}$ & *** \\ 
20 & 0.25 & 0.5 & $6.1 \times 10^{-8}$ & *** \\ 
20 & 0.25 & 1 & $3.4 \times 10^{-10}$ & *** \\ 
20 & 0.25 & 2 & $6.1 \times 10^{-8}$ & *** \\ 
20 & 0.25 & 4 & $1.2 \times 10^{-4}$ & *** \\ 
20 & 0.5 & 1 & $3.8 \times 10^{-1}$ &  \\ 
20 & 0.5 & 2 & $1.0 \times 10^{0}$ &  \\ 
20 & 0.5 & 4 & $1.1 \times 10^{-1}$ &  \\ 
20 & 1 & 2 & $3.9 \times 10^{-1}$ &  \\ 
20 & 1 & 4 & $1.5 \times 10^{-2}$ & * \\ 
20 & 2 & 4 & $1.1 \times 10^{-1}$ &  \\ 
40 & 0.25 & 0.5 & $1.7 \times 10^{-5}$ & *** \\ 
40 & 0.25 & 1 & $4.3 \times 10^{-5}$ & *** \\ 
40 & 0.25 & 2 & $5.4 \times 10^{-1}$ &  \\ 
40 & 0.25 & 4 & $2.1 \times 10^{-2}$ & * \\ 
40 & 0.5 & 1 & $8.3 \times 10^{-1}$ &  \\ 
40 & 0.5 & 2 & $2.6 \times 10^{-4}$ & *** \\ 
40 & 0.5 & 4 & $5.9 \times 10^{-11}$ & *** \\ 
40 & 1 & 2 & $5.3 \times 10^{-4}$ & ** \\ 
40 & 1 & 4 & $1.9 \times 10^{-10}$ & *** \\ 
40 & 2 & 4 & $3.2 \times 10^{-3}$ & ** \\ 
\bottomrule
\end{tabular}

\clearpage{}
\caption*{T-test of differences in estimated predictors by sample size}
\end{table}

\section{Discussion}\label{sec:discussion}

The results for Task A indicate that correlation perception is likely to be substantially influenced by the aspect ratio of the plot. The effects of this can be large - for example, for an aspect ratio of 1, underestimating a true correlation of 1 ($r$ 0.762) by approximately 0.75 leads to a perceived correlation around 0.25 ($r$ 0.245). Further, the number of polylines displayed (sample size) interacts with the aspect ratio - for low numbers of polylines an aspect ratio of $\geq$ 2 is much more likely to lead to accurate correlation estimation than an aspect ratio of 1. This difference appears to be mitigated by a larger sample size, with closely matched performance across AR 1, 2 and 4. This provides some support for the hypothesis that smaller between-line angles and length differences make it easier to estimate ‘closeness-to-parallel’ and reduces salience of line crossings, but where there are many more polylines the larger pool of potential comparisons reduces the difference. For the smallest aspect ratio (0.25), the interaction with sample size appears to have a much smaller effect, and plots at AR 0.25 are notably poor for estimating anything other than strong negative correlations, again appearing to support the increased angle-difference and line-length hypothesis unless the pattern formed is particularly distinctive, as is the case with the negative correlation ‘X’ pattern. Indeed, as Harrison et al. \cite{Harrison2014-ks} show, negative correlations are perceived much more strongly than positive in a PCP, so it is possible this effect counterbalances the reduced salience of the ‘X’ pattern for high aspect ratio.

\subsection{Asymmetric responses}
Positive correlations are harder to accurately perceive at low aspect ratios. For AR 0.25, responses to positive correlations are clustered around 0 (fig. \ref{fig:corr_results_bubble}), with fig. \ref{fig:corr_ob_accuracy} showing a substantial under-estimation of correlation strength for AR 0.25 and 0.5, indicating that for such small horizontal distances and higher between-line angles the pattern formed by positive correlations is hard to estimate. This effect is substantial, and is likely to lead to users completely missing all but the strongest positive correlations at low aspect ratios.
Where there is no correlation, responses for AR $\leq$  1 and AR 4 tend to skew negative or positive respectively, with a notably strong effect for AR 4 at a sample size of 160 (see fig. \ref{fig:corr_ob_accuracy} and fig. \ref{fig:corr_results_bubble} - right column, central value of Fisher-z 0). While Type I errors of this kind are problematic, in exploratory analysis tasks (where the goal is identifying \textit{potentially} interesting relationships) this may not present a significant downside and correlation could be confirmed by additional indicators. 

Much less asymmetry in response is shown for value tracing tasks, where performance is maximised at aspect ratio between 0.5 and 1. Decline in performance is most strongly apparent for high aspect ratio (4), particularly for sample size of 40. This fits with the hypothesis that the separation between axes makes lines hard to follow and it is plausible that the visual clutter introduced at the higher sample size makes the falloff in accuracy more acute. Potentially the reduced between-line angle may also lead to less readily apparent separation between lines with similar values. Similarly, for very low aspect ratio (0.25) the decline in accuracy is plausibly due to the difficulty of separating individual lines where the horizontal distance is very small.

Further work will be required to understand why higher sample sizes do not see a reduction (and potentially a small increase) in accuracy at AR 0.5 compared to 1, which is not apparent in the smaller sample sizes. Although this evaluation used static images, it is possible that even with brushing techniques to highlight the polyline of interest, some reduction in accuracy of rapid value estimation is possible for large aspect ratio ($>$4) due to separation distance of axes.

\subsection{Limitations}
The study was limited to two simple tasks with static images. As such, conclusions may not apply to interactive PCPs, where brushing and filtering interaction methods are available. Further, the height of the axes was set to ensure the full width of the PCP at AR 4 was visible with no scrolling required, meaning the \textit{area} between axes was not consistent between plots.

\section{Recommendations}
\paragraph{\textbf{Correlation estimation}}
Other methods should ideally be used in conjunction with a PCP \cite{Chang2018-jh}, but this is a common task and PCPs used in exploratory analysis may present distorted findings. As such, we suggest the following: \textbf{(i)} 
for low sample sizes, a more balanced estimation of positive and negative correlations may be achieved using an aspect ratio greater than 1; \textbf{(ii)} small aspect ratios are particularly poor for estimating positive correlations and should ideally be avoided unless used with a linked view; \textbf{(iii)} where sample size is unknown, an aspect ratio of 2 is likely to provide the most balanced option for correlation estimation.

\paragraph{\textbf{Value tracing}}
Although value tracing is frequently used in combination with interaction methods such as brushing, users performing exploratory tasks may use rapid non-interactive visual scanning to focus investigations and later interaction. 
Aspect ratios of between 0.5 and 1 are likely to be most effective to enable accurate value tracing. As such, we suggest the following: \textbf{(i)} very small ($<$0.25) and very large (4) aspect ratios strongly affect the performance of users in value tracing tasks and should be avoided, unless using brushing for fully-accurate value tracing, or enabling other interaction methods such as axis ordering and lensing techniques.

The above recommendations conflict if the PCP is intended (or is likely to be used) for both tasks. Primarily, a high aspect ratio providing balanced correlation estimation for low sample sizes will substantially reduce performance for value tracing. User information, interaction and linked views reduce this issue, although visualization designers will need to understand the primary use-case for the PCP.

\section{Future work}\label{sec:future}
Similar to the optimisation algorithms proposed for line charts (for example \cite{Heer2006-zm} and \cite{Talbot2011-ew}) or scatter plots (for example \cite{Fink2013-rl} or \cite{Micallef2017-ej}, we note that it may be possible to design an optimised custom spacing of axes. This would create an equally perceptually salient/identical discriminability of each z (i.e. varying the aspect ratio of the axes depending on a pre-calculated z), and could include adjustments for intended task.

The PCP is not necessarily ideally suited to correlation estimation tasks and supplementary visualizations or methods exist to mitigate some of the issues in correlation perception (such as 'angle uniform parallel coordinates' \cite{Zhang2023-iy} or 'indexed points parallel coordinates' \cite{Zhou2018-um}). However, these are not necessarily in common use and add both technical complexity and issues of unfamiliarity for users. Similarly, techniques exist to make it easier to trace values - for example  Kennedy and Graham \cite{Graham2004-jx} switch straight polylines to curves (evaluated by Holten and van Wijk \cite{Holten2010-np}). It is reasonable to assume that aspect ratio is likely to impact on some of these methods, although an evaluation of this is beyond the scope of this paper.

\section{Supplemental materials}
\label{sec:supplemental_materials}

All supplemental materials are available on Figshare at \url{https://dx.doi.org/10.6084/m9.figshare.31211107}.
% All supplemental materials are available on Figshare at \url{https://dx.doi.org/10.6084/m9.figshare.28869920}.

\begin{acks}
This work was supported in part by a grant from the UKRI Engineering and Physical Sciences Research Council.
\end{acks}

\bibliographystyle{SageV}

\end{document}